\documentclass[aps,prl,reprint,groupedaddress]{revtex4-2}

\usepackage{amsmath}
\usepackage{siunitx} 
\usepackage{amssymb}
\usepackage{graphicx}
\usepackage{bbm}
\usepackage{braket}
\usepackage[section]{placeins}
\usepackage{color}
\usepackage[normalem]{ulem}
\newcommand{\mr}{\mathrm}
\newcommand{\e}{\varepsilon}

\newcommand{\sectionprl}[1]{{\par\it #1.---}}

\begin{document}

\title{Microscopic origin of the effective spin-spin interaction in a 
semiconductor quantum dot ensemble}

\author{Frederik Vonhoff}
\author{Andreas Fischer}
\author{Kira Deltenre}
\author{Frithjof B.\ Anders}

\affiliation{Department of Physics, TU Dortmund University,
Otto-Hahn-Stra{\ss}e 4, 44227 Dortmund, Germany}
\date{\today}

\begin{abstract}
We present a microscopic model for a singly charged quantum dot (QD) ensemble
to reveal the origin of the long-range effective interaction between the electron spins in the QDs.
Wilson's numerical renormalization group (NRG) is used to calculate  
the magnitude and the spatial dependency of 
the effective spin-spin interaction mediated by the growth induced wetting layer.
Surprisingly, we found an antiferromagnetic Heisenberg coupling for very short inter-QD distances that is caused by
the significant particle-hole asymmetry of the wetting layer band at very low filling.
Using the NRG results obtained from realistic parameters as input for a semiclassical simulation 
for a large QD ensemble, we demonstrate that the experimentally reported phase shifts in the coherent spin dynamics
between single and two color laser pumping can be reproduced by our model, solving a longstanding mystery of the microscopic origin of the inter QD electron spin-spin interaction.
\end{abstract}

\maketitle

Spins confined in semiconductor QDs
have been discussed as candidates for implementation of quantum bits (qubits) in quantum information technologies \cite{GarcadeArquer2021,HansonSpinQdotsRMP2007}
since it allows integration into conventional semiconductor logic elements.
For information processing, qubit initialization and readout \cite{Elzerman04,Atatuere2006} are as important as manipulations of the spins. Optical control experiments in QD ensembles \cite{SpatzekGreilichBayer2011,Varwig2014}
as well as the measurements of the dephasing time as function of the 
laser spectral width \cite{FischerBayerAnders2018} in such samples
provided strong evidence for a long-range electron spin-spin interactions between the different
QDs in the ensemble of unknown microscopic origin. 
Initially, it was speculated \cite{SpatzekGreilichBayer2011} that it might be caused by an optically induced \cite{Optical_RKKY2002} Ruderman-Kittel-Kasuya-Yosida (RKKY) interaction \cite{RKKY1954,lit:Kasuya1956,lit:Yosida1957}. Since the laser pulse duration is of the order of ps such an optically induced interaction, however, 
would decay rather rapidly and is not compatible with the observed spin coherence on a scale of several ns.  

\begin{figure}[b]
\begin{center}
\includegraphics[width=0.3\textwidth]{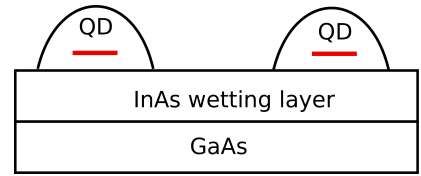}

\caption{Sketch of two quantum dots that are linked by the InAs wetting layer.}
\label{fig1-sketch}
\end{center}
\end{figure}

In order to make use of the intrinsic long range spin-spin interaction between the localized spins in QDs
by spin manipulation protocols, its origin needs to be understood.
In this letter, we propose a microscopic mechanism based on the analysis of a multi-impurity Anderson model \cite{Eickhoff2018,EickhoffMIAM2020,EickhoffKondoHole2021}. We start from a localized electron bound state in each QD 
that weakly hybridize with the conduction band (CB) of the thin wetting layer (WL) \cite{WL2002,Melnik_2003,WL2004} that is left below the QDs in the Stranski-Krastanow growth protocol. The basic setup is sketched in Fig.~\ref{fig1-sketch}. 
By including all virtual charge fluctuations in the leading order, significant corrections to the conventional textbook expression starting from an effective local moment picture have been reported \cite{ZitkoBonca2006,Eickhoff2018} for short distances. 
We believe that implementing backgates below the WL would allow to manipulate the properties of the low concentration WL electron gas and hence to control the effective spin-spin interaction.

Recently a growth procedure for InGaAs QD ensembles was proposed \cite{QDsnoWL2019} to
eliminate the WL. In such samples, the effective spin-spin interaction between the QD electron spins should be absent. 
While those types of QD ensembles
are produced to decrease the energy loss when used as photon emitters, a detailed investigation of their spin properties in electron doped samples is 
still missing to the best of our knowledge.

\sectionprl{Model}
The WL is treated as a free two-dimensional CB
\begin{eqnarray}
\label{eq:Hwl}
H_{\rm WL} &=& \sum_{\vec{k}} \frac{\hbar^2 \vec{k}^2}{2m^*}  c^\dagger_{\vec{k}\sigma} c_{\vec{k}\sigma}
\end{eqnarray}
in the isotropic effective mass approximation with $m^*/m_0= 0.023$ \cite{MadelingBook}.
The minimal model
for the $N_{\rm QD}$ QDs in the ensemble is given by
\begin{eqnarray}
H_{\rm QD} &=&  \sum_{i}^{N_{\rm QD}} \left( \sum_\sigma \e^d_{i} d^\dagger_{i\sigma} d_{i\sigma}
+U_i n^d_{i\uparrow} n^d_{i\downarrow} 
\right) ,
\end{eqnarray}
where $\e^d_{i} $ denotes the single particle energy of the bound electron state with spin $\sigma$ 
in the $i$-th QD and $U_i$ the corresponding Coulomb repulsion preventing the bound state to be doubly occupied. The creation (annihilation) operator $d^\dagger_{i\sigma} (d_{i\sigma}$) 
adds (removes) an electron to (from) the $i$-th QD. The electronic wave function is
localized but covers the whole diameter of the QD.
The parameters $U_i$ and $\e^d_i$ depend
 on the individual shape of the QD.
Using the experimental estimates \cite{NatureGoldhaberGordon1998}, we approximate $U_i\approx \SI{4}{\milli\electronvolt}$ for a QD with a diameter $D=\SI{25}{\nano\meter}$.

The concentration of the donors in the QD ensemble used in Ref.\ 
\cite{SpatzekGreilichBayer2011} was selected to match 
the QD density in the sample. Each QD can capture at least one electron if the bound state
$\e^d_i$ is below the CB of the WL. It remains empty when the chemical potential 
$\mu<\e^d_i$.
Spin spectroscopy experiments \cite{Glasenapp2016}, however, indicate that not all QDs are filled with one electron. This might be due to local imperfections that shift individual $\e^d_{i}$ to higher energies. 
Also doubly occupied QD states are possible when $\e^d_i < \mu-U_i$. 
Excluding this doubly occupied QD ground state configuration puts a lower bound on $\e^d_i$.
We assume that a fraction $q_{\rm WL}$
of the total donor excess electrons is filling up the WL such that the chemical potential $\mu>0$ lies within the CB of the WL. The upper bound of the chemical potential $\mu=\e_{F}$ is reached for $q_{\rm WL}=1$ yielding $\e_F\approx \SI{1}{\milli\electronvolt}$ for a dot density of \num{e10}cm$^{-2}$\cite{SpatzekGreilichBayer2011}.

The bound state of the $i$-th QD can tunnel with a finite tunneling amplitude $V^{i}_{m}$ into the Wannier orbital $m$ of the WL. The resulting hybridization term between the bound state of each QD and the WL takes the form
\begin{eqnarray}
H_{\rm hyb} &=& \sum_{i=1}^{N_{\rm QD}} \sum_\sigma \sum_{m} 
\left(V^{i}_{m} d^\dagger_{i\sigma} c_{m \sigma}  + h.c.\right),
\label{eq:H-hyp}
\end{eqnarray}
where $c_{m\sigma}$ is the annihilation operator of the Wannier orbital at site $\vec{R}_m$ of the 2D WL and whose spatial Fourier transform is $c_{\vec{k}\sigma}$. 
 The total Hamiltonian of the coupled QD problem is given by
$H = H_{\rm WL} + H_{\rm QD}  + H_{\rm hyb} $
which is just a realization of a multi-impurity problem \cite{EickhoffMIAM2020}.

The effect of the WL onto the dynamics of the localized QD states is determined by the
hybridization function matrix \cite{Eickhoff2018,EickhoffMIAM2020}
\begin{eqnarray}
\Delta_{ij}(z) &=& 
\sum_{lm} 
\frac{1}{N} \sum_{\vec{k}} \frac{ [V^{i}_l]^* V^j_m 
e^{i\vec{k} (\vec{R}_m-\vec{R}_l)}
}{z-\e_{\vec{k}}} .
\label{equ:Delta}
\end{eqnarray}
In the wide band limit, i.\ e.\ $V_0/D\ll 1$ where $D$ is the band width of the WL CB,
$t_{ij}=\Re \Delta_{ij}(\mu)$
generates an effective hopping between the QDs $i$ and $j$.
The averaged distance between the QDs is of the same order of magnitude as the Fermi wave length $\lambda_F$. The distance variations between the WL sites, however, are small compared to $\lambda_F$. 
Consequently, we can replace  $\vec{R}_m - \vec{R}_{l}$
by the distance between the two centers of the QDs, i.\ e.\ $\vec{R}_i - \vec{R}_j$,
and include the spatial extension of the QD 
\footnote{We checked that by an explicit calculation assuming a Gaussian distribution of $V^{i}_m$ centered at  $\vec{R}_j$ with a width given by the QD radius of $r_{\rm{QD}}\approx \SI{15}{\nano\meter}$.}
by defining $\bar V_i = \sum_m V^{i}_m$.
We introduce the average hybridization matrix element $V_0^2=\langle  \bar V_i^2
\rangle$ and define a reference energy scale $\Gamma_0 = \pi V_0^2\rho_0$, $\rho_0$ being the constant density of state
of the 2D WL CB. The charge fluctuation scale $\Gamma_0$ determines the order of magnitude of $\Delta_{ij}(z)$. 
The antiferromagnetic (AF) part of the RKKY interaction can be estimated \cite{Eickhoff2018,EickhoffMIAM2020} as 
$J_{\rm{AF},ij}^{\rm RKKY}\approx 4 t_{ij}^2/U\propto 4\Gamma_0^2/U$ serving as 
a first estimate for $\Gamma_0\approx 10-100\si{\micro\electronvolt}$.

In a conventional metal, the AF  part of the RKKY interaction is compensated by the FM contribution for $|\vec{R}|\to 0$ \cite{Jones_et_al_1987,ZitkoBonca2006}.
The relevant parameter regime of our model, however, does not fit the conventional regime discussed in the literature
where $D,U>\bar J_{K}$,  
$\bar J_{K}$ being the averaged local Heisenberg coupling $J_{K}$ between the QD electron spin and the local CB spin density \cite{FischerKleinRKKY2D1975}.
The energy $\e^d_i$ is located below the lower band edge of the CB. Many Wannier orbitals contribute to $\bar V_i$, and virtual charge fluctuations in the localized orbitals below the CB continuum are allowed. 
We expect deviations between the correct effective spin-spin interaction
within our model and the prediction from the conventional two stage mapping of the QD model: first onto a Kondo ensemble model by applying a Schrieffer-Wolff transformation (SWT) \cite{SchriefferWol66} that is $O(V^2_0)$ followed by a 
perturbative estimation of the $J^{\rm RKKY}_{ij}$ \cite{FischerKleinRKKY2D1975} in $O(V_0^4)$. 
Significant corrections 
to $J^{\rm RKKY}_{ij}$ have been already reported \cite{ZitkoBonca2006,Eickhoff2018} at short distances since additional terms in $O(V_0^4)$ neglected in the two stage mapping become important.

\sectionprl{Numerical renormalization group approach}
To circumvent the inconsistency problem of the two stage perturbative approach, we apply the 
numerical renormalization group (NRG) \cite{KrishWilWilson80a,BullaCostiPruschke2008} to the Hamiltonian with $N_{\rm QD}=2$ to determine 
the distance dependent effective Heisenberg interaction between the localized electron spins.  
For each distance $R=|\vec{R}_1-\vec{R}_2|$, the two QD problem is mapped onto an effective two band model \cite{Jones_et_al_1987,AffleckLudwigJones1995,Eickhoff2018}, where we used 
$\Gamma_0$, $\mu=\e_{F}$ and $\e_i^d = \e^d$ 
as the adjustable parameters
such that each QD remains singly charged. Note that the 
NRG mapping of the strong asymmetric bands onto a Wilson chain \cite{KrishWilWilson80a,BullaCostiPruschke2008} 
leads to modification of the hopping parameters \cite{RobertHager2007} and of the NRG fixed point (FP) spectrum
\footnote{See Supplemental Material for technical additional details.}.
This adjustment only depends on the support of the CB density of states and, therefore, is independent of the QD distance $R$.

\begin{figure}[t]
\begin{center}

\includegraphics[scale=1]{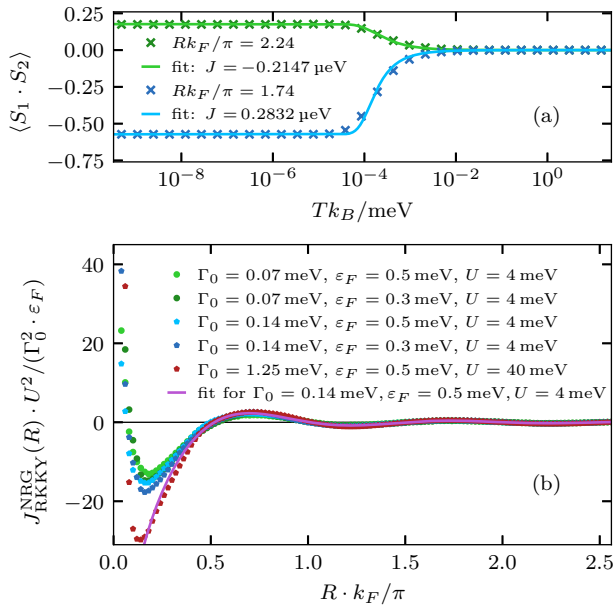}

\caption{NRG results: (a)  $\langle \vec{S}_1 \vec{S}_2 \rangle(T,R)$
vs $T$ for two distances $R$ and the fit to an interacting two spin model
for extracting $J$.
(b)  
$J^{\rm NRG}_{\rm RKKY}(R)/(\Gamma_0^2\e_{F})$ vs  $Rk_{F}/\pi$ 
for different $\Gamma_0$ and $\e_{F}$.
We added a fit to Eq.\ \eqref{eq:Jrkky} as solid magenta line. 
NRG parameters $\Lambda=5$, high energy cutoff $D=\SI{1}{\electronvolt}$, $\e^d = \SI{-1.5}{\milli\electronvolt},$ and $\bar \beta = 40$.
}
\label{fig-NRG-Jrkky}
\end{center}
\end{figure}

Using the NRG, we calculated the temperature dependent spin-spin correlation function 
$\langle \vec{S}_1 \vec{S}_2 \rangle(T,R)$
as shown for two distances in Fig.\ \ref{fig-NRG-Jrkky}(a). The sign of 
$\langle \vec{S}_1 \vec{S}_2 \rangle(T,R)$
determines the sign of  $J_{\rm RKKY}^{NRG}(R)$, and $|J_{\rm RKKY}^{NRG}(R)|$ is obtained from the fit to the universal functions of the spin-spin correlation function of a $s=1/2$ toy model, 
$S_{12}(T)= \mu_{\rm eff}^2  (e^{-\beta J}-1)/(1+3e^{-\beta J})$, where the parameter $\mu_{\rm eff}^2$ includes 
possible Kondo screening effects of the localized spins \footnote{$\mu_{\rm eff}^2=s(s+1)$ for unscreened spins.}.
We added the fit functions 
as solid lines to Fig.\ \ref{fig-NRG-Jrkky}(a).
The extracted $J_{\rm RKKY}^{NRG}(R)=J$ are shown in Fig.\ \ref{fig-NRG-Jrkky}(b)
for different fixed set of parameters.

For two localized moments coupled by a Heisenberg interaction $J_{K}$ 
to the local spin-density of 2D CB with a quadratic dispersion in 2D,
Klein and Fischer \cite{FischerKleinRKKY2D1975} derived the analytic expression 
\begin{eqnarray}
\label{eq:Jrkky}
J^{\rm RKKY}_{ij} &=& - 
\rho_0 \bar J^2_{K}  \frac{\bar v k_{F}^2}{4\pi}\\
&& \times
\left[J_0(k_{F} R) N_0(k_{F} R) + J_1(k_{F} R) N_1(k_{F} R) \right]
\nonumber
\end{eqnarray}
for the effective RKKY interaction,
where $J_l(x)(N_l(x))$ is the Bessel (Neumann) function of order $l$ and $\bar v$ the area of the 2D unit cell.
We added a fit to the expression Eq.\ \eqref{eq:Jrkky} to our NRG data as solid magenta line in Fig.\ \ref{fig-NRG-Jrkky}(b). 
The NRG results follow excellently the analytic predictions for larger distances. {The magnitude of $J_{\rm RKKY}^{NRG}(R)$
is proportional to $\Gamma_0^2\e_{F}$
as expected, but the absolut value differs
for those predicted by Eq.\ \eqref{eq:Jrkky}.
$J_{\rm RKKY}^{NRG}(R)$ remains invariant under rescaling $\alpha\Gamma_0,\alpha U,\alpha \e^d$ for $R k_{F}/\pi>0.5$. 
For short distances, $R k_{F}/\pi< 0.25$,
however, we observe significant deviations: 
The NRG reveals an AF RKKY interaction in contrary to the conventional RKKY result. The origin of this surprising effect can be
linked to the large effective single-particle hopping between the QD orbitals induced by the very strong particle-hole asymmetry of the CB for small WL fillings. The conventional two stages perturbation theory 
is inconsistent and requires additional corrections in $O(V_0^4)$ \cite{ZitkoBonca2006,Eickhoff2018}.

\sectionprl{Spin polarization in laser pulsed QD ensembles}
After establishing the distance dependent effective interaction between the electron spins in the different QDs,
we investigate its influence on an ensemble of singly charged QD in an external magnetic field 
subject to a two color laser pumping with circularly polarized light \cite{SpatzekGreilichBayer2011}.
We used a semiclassical simulation \cite{Glazov2012,Jaeschke2017} 
for the spin dynamics of a QD ensemble model \cite{FischerBayerAnders2018}
\begin{eqnarray}
H_{\rm array} &=& \sum_i^{N_{\mr{QD} }} H_1^{(i)} + \sum_{i<j} J_{ij} \vec{S}^{(i)}   \vec{S}^{(j)},
\end{eqnarray}
where the central spin model (CSM) $H_1^{(i)}$,
\begin{align}
	H_1^{(i)} =& g_{e}^{(i)} \mu_{\mr{B}} \vec{B}_{\mr{ext}} \vec{S}^{(i)} 
	+ \sum_{k=1}^{N_i} g_{N,k}^{(i)} \mu_\mr{N} \vec{B}_\mr{ext} \vec{I}_k^{(i)} 
	\nonumber\\
	 &
	+ \sum_{k=1}^{N_i} A_k^{(i)} \vec{I}_k^{(i)} \vec{S}^{(i)} \label{eq:H1},
\end{align}
includes the electron and nuclear Zeeman term in the external magnetic field $\vec{B}_{\mr{ext}}$
as well as the hyperfine coupling between the $N_i$ nuclei spins denoted by $\vec{I}_k^{(i)}$ and the electron spin $\vec{S}^{(i)}$
\cite{Gaudin1976,Merkulov2002,CoishLoss2004,ZangHarmon2006,ChenBalents2007}.
We generated random 2D ensembles of $N_{\mr{QD}}$ QDs 
with a dot density of \num{e10}cm$^{-2}$ \cite{SpatzekGreilichBayer2011}
and assigned $J_{ij}= x_ix_j J^{\rm NRG}_{\rm RKKY}(R,\e_F,\Gamma_0)$ 
from the NRG data depicted in Fig.\ \ref{fig-NRG-Jrkky}(b) 
where $x_i$ accounts for the variations of $\bar V_i$.
We attributed 
$N_{\rm QD}/2$ randomly selected QDs to be resonant to one of the two different laser frequencies and, therefore,
 also assigned 
the corresponding $g_{e}^{(i)}$. 
We set the average $\bar g_e=0.55$, the ratios between the two subsets $g_{e}^{1}/g^2_e=1.03$,
and $z=\bar g_e \mu_{B}/(\bar g_{N} \mu_{N}) = 1/800$.

We run three different types of semiclassical simulations for the spin dynamics with  $I_k=3/2$ nuclear spins: (i) a box model with $A_k=A_0=\mathrm{const.}$, $N_{\rm QD}=10000$, and $N_i=10^6$ nuclear spins,
(ii) a simulation for a frozen nuclear spin dynamics (FOA) \cite{Merkulov2002} for 
$N_{\rm QD}=10000$ 
and $N_i$ as in (i) and  (iii) an $A_k$ distribution with $N_{\mr{QD}}=1000$ and $N_i=1000$ 
including the full nuclear spin dynamics.  
In all cases, we kept 
the characteristic energy scale $T^* = (\sum_k (A_k^{(i)})^2\langle I^2_k\rangle)^{-1/2} =\SI{1}{\nano\second}$ fixed.
 The technical aspects of the simulation including the treatment of the laser pulse using a trion excitation cycle can be found in the literature \cite{Jaeschke2017,FischerBayerAnders2018} which is summarized in the Supplemental Material.

The experiments reported in Ref.\ \cite{SpatzekGreilichBayer2011} were performed in a transversal magnetic field of \SI{1}{\tesla}, and laser pulses of two different colors were applied pumping two different resonant QD subsets. In order to reproduce one of the experimental key results, we first determined a reference curve for $\langle S_z(t)\rangle$ of subset 1 subjected to a circular polarized laser pulse at $t=0$ that shows damped coherent oscillations characterized by the Larmor frequency. Then we run the same setup but apply a second circular polarized laser pulse with a color resonating with subset 2 at a moment when $\langle S_z(t)\rangle$ of subset 1 reached a minimum, i.\ e.\ at a delay of $\Delta t\approx 
\SI{0.111}{\nano\second}$.

\begin{figure}[t]
\begin{center}

\includegraphics[scale=1]{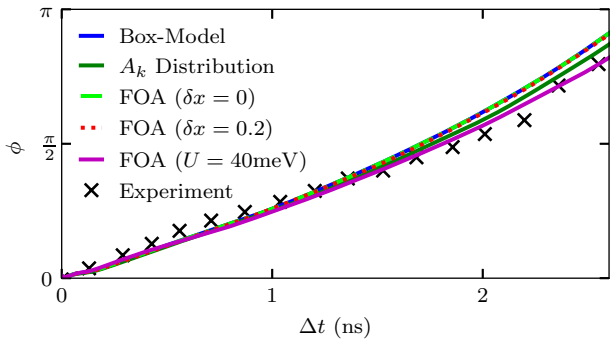}

\caption{Phase shift $\phi$ in the coherent electron spin oscillation between a single color pumping 
reference curve and a two color laser pumped QD ensemble.
The experimental data are taken from Ref.\ \cite{SpatzekGreilichBayer2011},
the simulations are for $A_k=\mathrm{const.}$ (box model), a frozen Overhauser field approximation (FOA)
including some disorder of the $x_j$ as well as a distribution of the $A_k$.
}
\label{fig-phase-shift}
\end{center}
\end{figure}

We tracked the relative phase shifts of the coherent oscillations between the reference curve 
 $\langle S_z(t)\rangle$ and those in the presence of a laser pulse onto subset 2. 
The sign of the phase shift depends only on the relative sign of the circular polarization. The result is plotted in Fig.~\ref{fig-phase-shift}.
As in the experiment, the phase shift increases roughly linearly in time. 
The slope is determined by overall magnitude of $J^{\rm RKKY}_{ij}(R)$. The results that resembles the experiments the best were obtained for $\Gamma_0= \SI{140}{\micro\electronvolt}$ and $\e_{F}= \SI{0.5}{\milli\electronvolt}$ corresponding to $q_{\rm WL}=0.5$ and $U/\Gamma_0=29$. 
Leaving the ratio  $U/\Gamma_0=29$ fixed, we can also reproduce the experimental phase shifts with 
$\Gamma_0= \SI{1.25}{\milli\electronvolt},\e^d=\SI{-15}{\milli\electronvolt}$, $U=\SI{40}{\milli\electronvolt}$ indicating the physics is driven by an effective lokal Kondo coupling \cite{SchriefferWol66} plus corrections in $O(V^4_0)$ for short distances.
Note that the individual hybridization strength 
between the bound electron QD state and the local Wannier orbital remains small and is of the order $V_m^i =10-40\,\si{\micro\electronvolt}$. Only by the summation to $\bar V_i$, a significant contribution arises.

We run the simulation for fixed $x_i=1$ and 
for the FOA  additional with a Gaussian normal distribution with $\delta x=0.2$ 
to reveal the  influence of the local derivation of the QD hybridization matrix element from its average value. 
The phase shifts are nearly independent of the distribution of local hybridization matrix elements $\bar V_i$. The major disorder contributions stems from the randomly distribution of distances to the neighbouring QDs. As a result, each QD senses a slightly different effective coupling converting the quadratic increase of the phase shift in a two QD model
to a linear increase  for short times.
Tracking the full dynamics of the Overhauser field is not required on the time scale of \SI{3}{\nano\second}: 
The frozen Overhauser approximation and the case $A_k^{(i)}=A_0$ yields the almost same results as the full nuclear spin dynamics but with a reduced $N_{{\rm QD}}$. We also run the simulation for a $J^{\rm RKKY}_{ij}(R)$
after rescaling  $\Gamma_0,U,\e^d$ with a factor of 10 to accommodate for errors in the rough parameter estimates. The slope of the phase shift $\phi$ came out about 50\% to large requiring a reduction of $\Gamma_0$. The origin of the increase is caused by contribution of the first minimum in 
$J^{\rm RKKY}_{ij}(R)$ at around $R K_F/\pi\approx 0.2$ where we observe 
significant deviations from the analytic prediction.
Overall, the simulation data agree remarkably well with the experimental findings even over a large range of parameters.

\sectionprl{Conclusion}
In this letter, we present a QD WL model to explain the microscopic origin of the inter QD electron spin-spin interaction conjectured in Ref.\ \cite{SpatzekGreilichBayer2011}. We used Wilson's NRG \cite{BullaCostiPruschke2008} to extract the strength and distance dependency of this 
spin-spin interaction mediated by the WL via an RKKY based mechanism. While the long distance behavior is in agreement with the low-filling perturbative RKKY result \cite{FischerKleinRKKY2D1975}, we found significant deviations for shorter distances. 
The unconventionally AF RKKY interaction at short distances can be connected to the large effective inter QD hopping between the QD orbitals \cite{Eickhoff2018,EickhoffMIAM2020}
as a consequence of the very low band filling. 
The conventional RKKY approach \cite{FischerKleinRKKY2D1975} starts already from an effective local moment picture and does not include all corrections in $O(V_0^4)$.

We used our NRG data as input for a semiclassical simulation of the spin dynamics in a coupled QD ensemble which is able 
to reproduce the experimental phase shifts reported for two-color pumping setup \cite{SpatzekGreilichBayer2011} very accurately 
for a variety of different realistic parameters. Although the interaction in the effective model might appear substantial, one has to bare in mind that all WL Wannier orbital below a QD  contribute in Eq.\ \eqref{eq:H-hyp} only by a very weak coupling.

Our theory solves the long standing mystery about the microscopic origin of the inter QD spin-spin interaction. We predict that this interaction can be eliminated by a QD growth process that excludes an InAs WL \cite{QDsnoWL2019}.
By gating the 2D WL,  it might be possible to modulate the 2D electron gas and, therefore, control the effective
Heisenberg interactions $J^{\rm RKKY}_{ij}(R)$ by a change of gate voltage.

\begin{acknowledgments}
The authors would like to thank fruitful discussions with Alex Greilich who also provided the experimental raw data of Ref.\ \cite{SpatzekGreilichBayer2011} added to Fig 2(b).
We acknowledge financial support by the Deutsche Forschungsgemeinschaft through the transregio TRR 160 within the Projects No. A4 and No. A7. 

\end{acknowledgments}
%

\end{document}